\documentclass{article}

\usepackage{microtype}
\usepackage{graphicx}
\usepackage{subfigure}
\usepackage{microtype}
\usepackage{graphicx}
\usepackage{subfigure}
\usepackage{booktabs} % for professional tables
\usepackage{amssymb}
\usepackage{stackengine}
\usepackage{amsmath}
\usepackage{amsfonts}
\usepackage{multirow}
\usepackage{todonotes}
\usepackage{mathrsfs}
\usepackage{booktabs} % for professional tables

\usepackage{hyperref}

\usepackage[accepted]{icml2019}

\icmltitlerunning{Imperceptible, Robust, and Targeted Adversarial Examples for Automatic Speech Recognition}

\begin{document}

\twocolumn[
\icmltitle{Imperceptible, Robust, and Targeted  \\Adversarial Examples for Automatic Speech Recognition}

\icmlsetsymbol{equal}{*}

\begin{icmlauthorlist}
\icmlauthor{Yao Qin}{to}
\icmlauthor{Nicholas Carlini}{goo}
\icmlauthor{Ian Goodfellow}{goo}
\icmlauthor{Garrison Cottrell}{to}
\icmlauthor{Colin Raffel}{goo}
\end{icmlauthorlist}

\icmlaffiliation{to}{Department of CSE, University of California, San Diego, USA}
\icmlaffiliation{goo}{Google Brain, USA}

\icmlcorrespondingauthor{Yao Qin}{yaq007@eng.ucsd.edu}
\icmlcorrespondingauthor{Colin Raffel}{craffel@google.com}

% You may provide any keywords that you
% find helpful for describing your paper; these are used to populate
% the "keywords" metadata in the PDF but will not be shown in the document
\icmlkeywords{Machine Learning, ICML}

\vskip 0.3in
]

\printAffiliationsAndNotice{}  % leave blank if no need to mention equal contribution
%\printAffiliationsAndNotice{\icmlEqualContribution} % otherwise use the standard text.

\begin{abstract}
Adversarial examples are inputs to machine learning models designed by an adversary to cause an incorrect output.
So far, adversarial examples have been studied most extensively in the image domain.
In this domain, adversarial examples 
can be constructed by imperceptibly modifying images to cause misclassification,
and are practical in the physical world.
In contrast, current targeted adversarial examples applied to speech recognition systems
have neither of these properties: humans can 
easily identify the adversarial perturbations,
and they are not effective when played over-the-air.
This paper makes advances on both of these fronts.
First, we develop \emph{effectively imperceptible} audio adversarial examples
(verified through a human study)
by leveraging the psychoacoustic principle of auditory masking, 
while retaining $100\%$ targeted success rate
on arbitrary full-sentence targets.
Next, we make progress towards physical-world over-the-air audio adversarial examples by
constructing perturbations which remain effective even after applying
realistic simulated
environmental distortions.
\end{abstract}
\section{Introduction}
Adversarial examples \cite{szegedy2013intriguing} are inputs that have been 
specifically designed by an adversary to
cause a machine learning algorithm to produce a misclassification \cite{biggio2013evasion}.
Initial work on adversarial examples focused mainly on the domain of
image classification.
In order to differentiate properties of adversarial examples on neural networks in
general from properties which hold true only on images, it is important to
study adversarial examples in different domains.
Indeed, adversarial examples are known to exist on domains ranging from reinforcement
learning \cite{huang2017adversarial} to reading comprehension \cite{jia2017adversarial} to speech recognition \cite{Carlini2018AudioAE}. 
This paper focuses on the latter of these domains, where \cite{Carlini2018AudioAE}
showed that any given source audio sample can be perturbed
slightly so that an automatic speech recognition (ASR) system
would transcribe the audio as any different target sentence.

To date, adversarial examples on ASR
differ from adversarial examples on images in two key ways.
First,
adversarial examples on images are imperceptible to humans:
it is possible to generate an adversarial
example without changing the 8-bit brightness representation \cite{szegedy2013intriguing}.
Conversely, adversarial examples on ASR systems are
often perceptible. While the perturbation introduced is often small in magnitude, upon listening it is obvious that the added perturbation
is present \cite{schonherr2018adversarial}.
Second, adversarial examples on images work in the physical world \cite{kurakin2016adversarial}
(e.g., even when taking a picture of them). In contrast, adversarial examples
on ASR systems do not yet work in such an ``over-the-air'' setting where
they are played by a speaker and recorded by a microphone.

In this paper\footnote{The project webpage is at \small{ \url{http://cseweb.ucsd.edu/~yaq007/imperceptible-robust-adv.html}}}, we improve the construction of adversarial examples on the ASR system
and match the power of attacks on images by developing adversarial examples which are imperceptible, and make steps towards robust adversarial examples.

In order to generate imperceptible adversarial examples, we depart from the common
$\ell_p$ distance measure widely used for adversarial example research.
Instead, we make use of the psychoacoustic principle of auditory masking,
and only add the adversarial perturbation to regions of the audio where it will not be heard by a human,
even if this perturbation is not ``quiet'' in terms of absolute energy.

Further investigating properties of adversarial examples which appear to be different from images, we examine the ability of an adversary to construct physical-world
adversarial examples~\cite{kurakin2016adversarial}.
These are inputs that, even after taking into account the distortions introduced by
the physical world, remain adversarial upon classification.
We make initial steps towards developing audio which
can be played over-the-air 
by designing audio which remains adversarial after being
processed by random room-environment simulators~\cite{scheibler2018pyroomacoustics}.

Finally, we additionally demonstrate that our attack is capable of attacking a modern, state-of-the-art Lingvo ASR system~\cite{shen2019lingvo}.

\section{Related Work}

We build on a long line of work studying the robustness of neural networks.
This research area largely began
with \cite{biggio2013evasion,szegedy2013intriguing}, who first studied
\emph{adversarial examples} for deep neural networks.

This paper focuses on adversarial examples on automatic speech recognition
systems. Early work in this space
\cite{gong2017crafting,cisse2017houdini} was successful when generating
\emph{untargeted} adversarial examples that produced incorrect, but arbitrary, transcriptions.
A concurrent line of work succeeded at generating targeted attacks in practice, even when
 played over a speaker and recorded by a microphone (a so-called ``over-the-air''
attack) but only by both (a) synthesizing
completely new audio
and (b) targeting older, traditional (i.e., not neural network based) speech recognition systems \cite{carlini2016hidden,zhang2017dolphinattack,song2017inaudible}.

These two lines of work were partially unified by \citet{Carlini2018AudioAE} who constructed
adversarial perturbations for speech recognition systems targeting arbitrary 
(multi-word) sentences. However, this attack was neither effective over-the-air,
nor was the adversarial perturbation completely inaudible; while the perturbations it
introduces are very quiet, they can be heard by a human (see \S~\ref{sec:humanstudy}).
Concurrently, the CommanderSong \cite{yuan2018commandersong} attack developed adversarial examples that are effective
over-the-air but at a cost of introducing a significant perturbation to the original audio.
 
Following this, concurrent work with ours develops
attacks on deep learning ASR systems that either work over-the-air or 
are less obviously perceptible.
\begin{itemize}
\item \citet{yakura2018robust},
create adversarial examples which can be played over-the-air. These attacks are
highly effective on short two- or three-word phrases, but not on the full-sentence phrases
originally studied. Further, these adversarial examples often have a significantly
larger perturbation, and in one case, the perturbation introduced had a \emph{higher} magnitude
than the original audio.

\item \citet{schonherr2018adversarial}
work towards developing attacks that are less perceptible
through using ``Psychoacoustic Hiding'' and attack the Kaldi system,
which is partially based on neural networks but also uses some ``traditional'' 
components such as a Hidden Markov Model instead of an RNN for final classification.
Because of the system differences we can not directly compare our results
to that of this paper, but we encourage the reader to listen to examples
from both papers.
\end{itemize}

Our concurrent work manages to achieve both of these results (almost) simultaneously: we generate adversarial examples that are both nearly
imperceptible and also remain effective after simulated distortions.
Simultaneously, we target a state-of-the-art network-based ASR system, Lingvo, as opposed to Kaldi and generate full-sentence adversarial
examples as opposed to targeting short phrases.

A final line of work extends adversarial example generation on ASR systems
from the white-box setting (where the adversary has complete knowledge of the underlying
classifier) to the black-box setting 
\cite{khare2018adversarial,taori2018targeted} (where the adversary is only allowed to query
the system).
This work is complementary and independent of ours: we assume a white-box threat model.

\section{Background}
\subsection{Problem Definition}
Given an input audio waveform $x$, a target transcription $y$ and an automatic
speech recognition (ASR) system $f(\cdot)$ which outputs a final transcription,
our objective is to construct an imperceptible and targeted adversarial example $x'$
that can attack the ASR system when played over-the-air. That is, we seek to find
a small perturbation $\delta$, which enables $x' = x+\delta$ to meet three requirements: 

\begin{itemize}
    \item \textbf{Targeted}: the classifier is fooled so that $f(x') = y$ and $f(x)\neq y$.
    Untargeted adversarial examples on ASR systems often only introduce spelling errors and
    so are less interesting to study.
    
    \item \textbf{Imperceptible}: $x'$ sounds so similar to $x$ that humans cannot differentiate $x'$ and $x$ when listening to them.
    \item \textbf{Robust}: $x'$ is still effective when played by a speaker and recorded by a microphone in an over-the-air attack. (We do not achieve
    this goal completely, but do succeed at simulated environments.)
\end{itemize}

\subsubsection{ASR Model}
We mount our attacks on the \textbf{Lingvo} classifier~\cite{shen2019lingvo}, a state-of-the-art sequence-to-sequence model \citep{sutskever2014sequence} with attention \citep{bahdanau2014neural} whose architecture is based on the Listen, Attend and Spell model~\cite{chan2016listen}.
It feeds filter bank spectra into an encoder consisting of a stack of convolutional and LSTM layers, which conditions an LSTM decoder that outputs the transcription.
The use of the sequence-to-sequence framework allows the entire model to be trained end-to-end with the standard cross-entropy loss function.

\subsubsection{Threat Model}
In this paper, as is done in most prior work, 
we consider the white box threat model where the adversary has
full access to the model as well as its parameters.
In particular, the adversary is allowed to compute gradients through the
model in order to generate adversarial examples.

When we mount over-the-air attacks, we do not assume we know the exact
configurations of the room in which the attack will be performed.
Instead, we assume we know the \emph{distribution} from which the room
will be drawn, and generate adversarial examples so as to be effective on
any room drawn from this distribution.

\subsection{Adversarial Example Generation}
Adversarial examples are typically generated by performing gradient descent
with respect to the input on a loss
function designed to be minimized when the input is adversarial \cite{szegedy2013intriguing}.
Specifically, let $x$ be an input to a neural network $f(\cdot)$, let $\delta$ be a
perturbation, and let $\ell(f(x),y)$ be a loss function that is minimized when $f(x) = y$.
Most work on adversarial examples focuses on minimizing the max-norm
($\lVert{}\cdot\rVert{}_\infty$ norm) of $\delta$.
Then, the typical adversarial example generation algorithm \cite{szegedy2013intriguing,carlini2017towards,madry2017towards} solves
\vspace{-0.1mm}
\begin{align*}
& \text{minimize} \,\,\,\ell(f(x+\delta),y) + \alpha \cdot \lVert{}\delta\rVert{} \\
    & \text{such that} \,\,\,\lVert{}\delta\rVert{} < \epsilon
\end{align*}
\vspace{-1mm}
(where in some formulations $\alpha=0$). Here, $\epsilon$ controls the maximum
perturbation introduced.

To generate adversarial examples on ASR systems, \citet{Carlini2018AudioAE} set
$\ell$ to the CTC-loss and use the max-norm which has the effect of adding
a small amount of adversarial perturbation consistently throughout the audio sample.

\section{Imperceptible Adversarial Examples}\label{imperceptible}

Unlike on images, where minimizing $\ell_p$ distortion between an image and the nearest
misclassified example yields a visually indistinguishable image, on audio, this is not
the case~\cite{schonherr2018adversarial}.
Thus, in this work, we depart from the $\ell_p$ distortion measures and instead 
rely on the extensive work which has been done in the audio space for capturing the
human perceptibility of audio.

\subsection{Psychoacoustic Models}
A good understanding of the human auditory system is critical in order to be able to construct imperceptible adversarial examples. In this paper, we use \textit{frequency masking}, which refers to the phenomenon that a louder signal (the ``masker'') can make other signals at nearby frequencies (the ``maskees'') imperceptible~\cite{mitchell2004introduction, lin2015principles}. In simple terms, the masker can be seen as creating a ``masking threshold'' in the frequency domain. Any signals which fall under this threshold are effectively imperceptible.

 Because the masking threshold is measured in the frequency domain, and because audio signals change rapidly over time, we first compute the short-time Fourier transform of the raw audio signal to obtain the spectrum of overlapping sections (called ``windows'') of a signal. The window size $N$ is 2048 samples which are extracted with a ``hop size'' of 512 samples and are windowed with the modified Hann window. We denote $s_x(k)$ as the $k$th bin of the spectrum of frame $x$.

Then, we compute the log-magnitude power spectral density (PSD) as follows:
\vspace{-3mm}
\begin{equation}\label{p}
    p_x(k) = 10 \log_{10} \left| \frac{1}{N} s_x(k)\right|^2.
    \vspace{-1mm}
\end{equation}
The normalized PSD estimate $\bar{p}_x(k)$ is defined by~\citet{lin2015principles}
\begin{equation}\label{pd}
    \bar{p}_x(k) = 96 - \max_k\{ p_x(k) \} + p_x(k)
\end{equation}

\paragraph{Masking Threshold} Given an audio input, in order to compute its masking threshold, first we need to identify the maskers, whose normalized PSD estimate $\bar{p}_x(k)$ must satisfy three criteria: 1) they must be local maxima in the spectrum; 2) they must be higher than the threshold in quiet; and 3) they have the largest amplitude within 0.5 Bark (a psychoacoustically-motivated frequency scale) around the masker's frequency. Then, each masker's masking threshold can be approximated using the simple two-slope spread function, which is derived to mimic the excitation patterns of maskers. Finally, the global masking threshold $\theta_x(k)$ is a combination of the individual masking threshold as well as the threshold in quiet via addition (because the effect of masking is additive in the logarithmic domain). We refer interested readers to our appendix  and~\cite{lin2015principles} for specifics on computing the masking threshold. 

When we add the perturbation $\delta$ to the audio input $x$, if the normalized PSD estimate of the perturbation $\bar{p}_{\delta}(k)$ is under the frequency masking threshold of the original audio $\theta_x(k)$, the perturbation will be masked out by the raw audio and therefore be inaudible to humans. The normalized PSD estimate of the perturbation $\bar{p}_{\delta}(k)$ can be calculated via:
\begin{equation}\label{pdelta}
    \bar{p}_{\delta}(k) = 96 - \max_k\{p_x(k)\} + p_{\delta}(k).
\end{equation}
where $p_{\delta}(k) = 10\log_{10}|\frac{1}{N}s_{\delta}(k)|^2$ and $p_x(k) = 10 \log_{10} | \frac{1}{N} s_{x}(k)|^2$ are the PSD estimate of the perturbation and the original audio input.
\subsection{Optimization with Masking Threshold}
\paragraph{Loss function} Given an audio example $x$ and a target phrase $y$, we formulate the problem of constructing an imperceptible adversarial example $x' = x+\delta$ as minimizing the loss function $\ell(x, \delta, y)$, which is defined as:
\begin{equation}\label{loss}
    \ell(x, \delta, y) = \ell_{net}(f(x+\delta), y) + \alpha \cdot \ell_{\theta}(x, \delta) 
\end{equation}
where $\ell_{net}$ requires that the adversarial examples fool the audio recognition system into making a targeted prediction $y$, where $f(x) \neq y$. %In the DeepSpeech network~\cite{Hannun2014DeepSS}, the CTC loss function is used as $\ell_{net}$ while the simple cross entropy loss function is used in the Lingvo architecture~\cite{lingvo}.
In the Lingvo model, the simple cross entropy loss function is used for $\ell_{net}$.
The term $\ell_{\theta}$ constrains the normalized PSD estimation of the perturbation $\bar{p}_{\delta}(k)$ to be under the frequency masking threshold of the original audio $\theta_x(k)$. The hinge loss is used here to compute the loss for masking threshold:
\begin{equation}\label{th}
    \ell_{\theta}(x, \delta) = \frac{1}{\lfloor \frac{N}{2} \rfloor + 1} \sum_{k=0}^{\lfloor \frac{N}{2} \rfloor} \max\big\{\bar{p}_{\delta}(k) - \theta_{x}(k), 0\big\},
\end{equation}
where $N$ is the predefined window size and $\lfloor x \rfloor$ outputs the greatest integer no larger than $x$. The adaptive parameter $\alpha$ is to balance the relative importance of these two criteria.

\subsubsection{Two Stage Attack}
Empirically, we find it is difficult to directly minimize the masking threshold loss function via backpropagation without any constraint on the magnitude of the perturbation $\delta$. This is reasonable because it is very challenging to fool the neural network and in the meanwhile, limit a very large perturbation to be under the masking threshold in the frequency domain.
In contrast, if the perturbation $\delta$ is relatively small in magnitude, then it will be much easier to push the remaining distortion under the frequency masking threshold.

Therefore, we divide the optimization into two stages: the first stage of optimization focuses on finding a relatively small perturbation to fool the network (as was done in prior work \cite{Carlini2018AudioAE}) and the second stage makes the adversarial examples imperceptible.

In the first stage, we set $\alpha$ in Eqn~\ref{loss} to be zero and clip the perturbation to be within a relatively small range. As a result, the first stage solves:
\begin{equation} \label{lnet}
\begin{split}
 & \text{minimize} \,\,\,\ell_{net}(f(x+\delta),y) \\
& \text{such that} \,\,\,\lVert{}\delta\rVert{} < \epsilon
\end{split}
\end{equation}
where $\lVert{}\delta\rVert{}$ represents the $\lVert{}\cdot\rVert{}_\infty$ max-norm of $\delta$. Specifically, we begin by setting $\delta=0$ and then on each iteration:
\begin{equation}
\delta \leftarrow \textnormal{clip}_\epsilon (\delta - lr_1 \cdot \textnormal{sign} (\nabla_\delta \ell_{net}(f(x+\delta), y))),
\end{equation} 
where $lr_1$ is the learning rate and $\nabla_\delta\ell_{net}$ is the gradient of $\ell_{net}$ with respect to $\delta$. We initially set $\epsilon$
to a large value and then gradually reduce it during optimization following~\citet{Carlini2018AudioAE}.

The second stage focuses on making the adversarial examples imperceptible, with an \emph{unbounded} max-norm; in this stage, $\delta$ is only constrained by the masking threshold constraints. Specifically, initialize $\delta$ with $\delta_{im}^*$ optimized in the first stage and then on each iteration:
\vspace{-0.5mm}
\begin{equation}
\delta \leftarrow \delta - lr_{2} \cdot \nabla_\delta \ell(x, \delta, y), 
\end{equation}
\vspace{-0.5mm}
where $lr_2$ is the learning rate and $\nabla_\delta\ell$ is the gradient of $\ell$ with respect to $\delta$. The loss function $\ell(x, \delta, y)$ is defined in Eqn.~\ref{loss}. The parameter $\alpha$ that balances the network loss $\ell_{net}(f(x+\delta), y)$ and the imperceptibility loss $\ell_\theta(x, y)$ is initialized with a small value, \emph{e.g.,} 0.05, and is adaptively updated according to the performance of the attack. Specifically, every twenty iterations, if the current adversarial example successfully fools the ASR system (i.e.\ $f(x+\delta) = y$), then $\alpha$ is increased to attempt to make the adversarial example less perceptible. Correspondingly, every fifty iterations, if the current adversarial example fails to make the targeted prediction, we decrease $\alpha$. We check for attack failure less frequently than success (fifty vs. twenty iterations) to allow more iterations for the network to converge. The details of the optimization algorithm are further explained in the appendix.

\vspace{-3mm}
\section{Robust Adversarial Examples}\label{secrobust}
\subsection{Acoustic Room Simulator}
In order to improve the robustness of adversarial examples when playing over-the-air, we use an acoustic room simulator to create artificial utterances (speech with reverberations) that mimic playing the audio over-the-air. The transformation function in the acoustic room simulator, denoted as $t$, takes the clean audio $x$ as an input and outputs the simulated speech with reverberation $t(x)$. First, the room simulator applies the classic Image Source Method introduced in~\cite{allen1979image,scheibler2018pyroomacoustics} to create the room impulse response $r$ based on the room configurations (the room dimension, source audio and target microphone's location, and reverberation time). Then, the generated room impulse response $r$ is convolved with the clean audio to create the speech with reverberation, to obtain $t(x) = x * r$ where $*$ denotes the convolution operation. To make the generated adversarial examples robust to various environments, multiple room impulse responses $r$ are used. Therefore, the transformation function $t$ follows a chosen distribution $\mathcal{T}$ over different room configurations. 
\vspace{-1.5mm}
\subsection{Optimization with Reverberations}
In this section, our objective is to make the perturbed speech with reverberation (rather than the clean audio) fool the ASR system. As a result, the generated adversarial examples $x' = x+\delta$ will be passed through the room simulator first to create the simulated speech with reverberation $t(x')$, mimicking playing the adversarial examples over-the-air, and then the simulated $t(x')$ will be fed as the new input to fool the ASR system, aiming at $f(t(x')) = y$. Simultaneously, the adversarial perturbation $\delta$ should be relatively small in order not to be audible to humans.

In the same manner as the Expectation over Transformation in~\cite{Athalye2018SynthesizingRA}, we optimize the expectation of the loss function over different transformations $t\sim \mathcal{T}$ as follows:
\begin{equation} 
\begin{split}
 & \text{minimize} \,\,\,\ell(x, \delta, y) = \mathop{\mathbb{E}}\limits_{t \sim \mathcal{T}} \big[\ell_{net}(f(t(x+\delta)), y)\big]\\
& \text{such that} \,\,\,\lVert{}\delta\rVert{} < \epsilon.
\end{split}
\end{equation}
Rather than directly targeting $f(x + \delta) = y$, we apply the loss function $l_{net}$ (the cross entropy loss in the Lingvo network) to the classification of the transformed speech $f(t(x + \delta)) = y$. We approximate the gradient of the expected value via independently sampling a transformation $t$ from the distribution $\mathcal{T}$ at each gradient descent step. 

In the first $I_{r_1}$ iterations, we initialize $\epsilon$ with a sufficiently large value and gradually reduce it following~\citet{Carlini2018AudioAE}. We consider the adversarial example successful if it successfully fools the ASR system under a single random room configuration; that is, if $f(t(x+\delta))=y$ for just one $t(\cdot)$.
Once this optimization is complete, we obtain the max-norm bound for $\delta$, denoted as $\epsilon^*_r$. We will then use the perturbation $\delta^*_r$ as an initialization for $\delta$ in the next stage.

Then in the following $I_{r_2}$ iterations, we finetune the perturbation $\delta$ with a much smaller learning rate. The max-norm bound $\epsilon$ is increased to $\epsilon_r^{**} = \epsilon^*_r + \Delta$, where $\Delta > 0$, and held constant during optimization. 
During this phase, we only consider the attack successful if the adversarial example successfully fools a set of randomly chosen transformations $\Omega = \{t_1, t_2, \cdots, t_M\} $, where $t_i \sim \mathcal{T}$ and $M$ is the size of the set $\Omega$. The transformation set $\Omega$ is randomly sampled from the distribution $\mathcal{T}$ at each gradient descent step. In other words, the adversarial example $x' = x + \delta$ generated in this stage satisfies $\forall t_i \in \Omega$, $f(t_i(x+\delta)) = y$. In this way, we can generate robust adversarial examples that successfully attack ASR systems when the exact room environment is not known ahead of time, whose configuration is drawn from a pre-defined distribution. More details of the algorithm are shown in the appendix.

It should be emphasized that there is a tradeoff between imperceptibility and robustness (as we will show experimentally in Section~\ref{sec:humanstudy}). If we increase the maximal amplitude of the perturbation $\epsilon_r^{**}$, the robustness can always be further improved. Correspondingly, it becomes much easier for humans to perceive the adversarial perturbation and alert the ASR system. In order to keep these adversarial examples mostly imperceptible, we therefore limit the $\ell_\infty$ amplitude of the perturbation to be in a reasonable range.

\section{Imperceptible and Robust Attacks}
By combining both of the techniques we developed earlier, we now develop
an approach to generate both imperceptible and robust adversarial examples.
This can be achieved by minimizing the loss
\begin{equation}
\ell(x, \delta, y) = \mathop{\mathbb{E}}\limits_{t \sim \mathcal{T}} \big[\ell_{net}(f(t(x+\delta)), y) + \alpha \cdot \ell_{\theta}(x, \delta)\big], 
\end{equation}
where the cross entropy loss function $\ell_{net}(\cdot)$ is again the loss used for Lingvo, and the imperceptibility loss $\ell_{\theta}(\cdot)$ is the same as that defined in Eqn~\ref{th}. Since we need to fool the ASR system when the speech is played under a random room configuration, the cross entropy loss $\ell_{net} (f(t(x + \delta)), y)$ forces the transcription of the transformed adversarial example $t(x + \delta)$ to be $y$ (again, as done earlier).

To further improve these adversarial examples to be imperceptible, we optimize $\ell_{\theta}(x, \delta)$ to constrain the perturbation $\delta$ to fall under the masking threshold of the clean audio in the  frequency domain. This is much easier compared to optimizing the hinge loss $\ell_{\theta}(t(x), t(\delta)) = \max\{\bar{p}_{t(\delta)}(k) - \theta_{t(x)}(k), 0\}$ because the frequency masking threshold of the clean audio $\theta_x(k)$ can be pre-computed while the masking threshold of the speech with reverberation $\theta_{t(x)}(k)$ varies with the room reverberation $r$. In addition, optimizing $\ell_{\theta}(x, \delta)$ and $\ell_{\theta}(t(x), t(\delta))$ have similar effects based on the convolution theorem that the Fourier transform of a convolution of two signals is the pointwise product of their Fourier transforms. Note that the speech with reverberation $t(x)$ is a convolution of the clean audio $x$ and a simulated room reverberation $r$, hence:
\begin{equation}\label{fourier}
\mathscr{F}\{t(x)\} = \mathscr{F}\{x * r\} = \mathscr{F}\{x\} \cdot \mathscr{F}\{r\}
\end{equation}
where $\mathscr{F}$ is the Fourier transform, $*$ denotes the convolution operation and $\cdot$ represents the pointwise product. We apply the short-time Fourier transform to the perturbation and the raw audio signal first in order to compute the power spectral density $\bar{p}_{t(\delta)}$ and the masking threshold $\theta_{t(x)}$ in the frequency domain. Since most of the energy in the room impulse response falls within the spectral analysis window size, the convolution theorem in Eqn~\ref{fourier} is approximately satisfied. Therefore, we arrive at:
\begin{equation}
(\bar{p}_{t(\delta)} - \theta_{t(x)}) \approx (\bar{p}_{\delta} - \theta_{x}) \cdot \mathscr{F}\{r\}.
\end{equation}

As a result, simply optimizing the imperceptibility loss $\ell_{\theta}(x, \delta)$ can help in finding the optimal $\delta$ and in constructing the imperceptible adversarial examples that can attack the ASR systems in the physical world.

Specifically, we will first initialize $\delta$ with the perturbation $\delta^{**}_r$ that enables the adversarial examples to be robust in Section~\ref{secrobust}. Then in each iteration, we randomly sample a transformation $t$ from the distribution $\mathcal{T}$ and update $\delta$ according to:
\begin{equation}
\delta \leftarrow \delta - lr_{3} \cdot \nabla_\delta \big[\ell_{net}(f(t(x +\delta), y)) + \alpha \cdot \ell_\theta(x, \delta))\big], 
\end{equation}
where $lr_3$ is the learning rate and $\alpha$, a parameter that balances the importance of the robustness and the imperceptibility, is adaptively changed based on the performance of adversarial examples. Specifically, if the constructed adversarial example can successfully attack a set of randomly chosen transformations, then $\alpha$ will be increased to focus more on imperceptibility loss. Otherwise, $\alpha$ is decreased to make the attack more robust to multiple room environments. The implementation details are illustrated in the appendix.

\section{Evaluation}
\subsection{Datasets and Evaluation Metrics}
\paragraph{Datasets} We use the LibriSpeech dataset~\cite{panayotov2015librispeech} in our experiments, which is a corpus of 16KHz English speech derived from audiobooks and is used to train the Lingvo system~\cite{shen2019lingvo}. We randomly select 1000 audio examples as source examples, and 1000 separate transcriptions from the test-clean dataset to be the targeted transcriptions. We ensure that each target transcription is around the same length as the original transcription because it is unrealistic and overly challenging to perturb a short audio clip (\emph{e.g.,} 10 words) to have a much longer transcription (\emph{e.g.,} 20 words). Examples of the original and targeted transcriptions are available in the appendix.

\vspace{-2mm}
\begin{table*}
\vskip 0.15in
\begin{minipage}{0.35\textwidth}
\begin{center}
\begin{tabular}{lcc}
\toprule
\textbf{Input} & Clean & Adversarial \\
\midrule
\textbf{Accuracy} (\%) & 58.60 & 100.00 \\
\textbf{WER} (\%) & 4.47 & 0.00 \\
\bottomrule
\end{tabular}
\end{center}
\caption{Sentence-level accuracy and WER for 1000 clean and (imperceptible) adversarially perturbed examples, fed without over-the-air simulation into the Lingvo model. In ``Clean'', the ground truth is the original transcription. In``Adversarial'', the ground truth is the targeted transcription. \label{non_ota_results}}
\end{minipage}\hfill
\begin{minipage}{0.62\textwidth}
\begin{center}
\begin{tabular}{lccccc}
\toprule
\textbf{Input} & Clean & \shortstack{Robust \\ ($\Delta = 300$)} & \shortstack{Robust \\ ($\Delta = 400$)} & \shortstack{Imperceptible \\ \& Robust} \\
\midrule
\textbf{Accuracy} (\%) & 31.37 & 62.96 & 64.64 & 49.65 \\
\textbf{WER} (\%) & 15.42 & 14.45 & 13.83 & 22.98 \\
\bottomrule
\end{tabular}
\end{center}
\caption{Sentence-level accuracy and WER for 100 clean and adversarially perturbed examples, fed with over-the-air simulation into the Lingvo model. The ground truth for ``clean'' inputs is the original transcription while the ground truth is the targeted transcription for the adversarial inputs. The perturbation is bounded by $\lVert{\delta\rVert{}} < \epsilon^*_r + \Delta$. \label{ota_results}}
\end{minipage}
\vspace{-4mm}
\end{table*}

\paragraph{Evaluation Metrics} For automatic speech recognition, we evaluate our model using the standard word error rate (WER) metric, which is defined as $\textnormal{WER} = \frac{S + D + I}{N_W} \times 100\%$, where $S$, $D$ and $I$ are the number of substitutions, deletions and insertions of words respectively, and $N_W$ is the total number of words in the reference. 

We also calculate the success rate (sentence-level accuracy) as $\textnormal{Accuracy} = \frac{N_s}{N_a} \times 100\%$, where $N_a$ is the number of audio examples that we test, and $N_s$ is the number of audio examples that are correctly transcribed. Here, ``correctly transcribed'' means the original transcription for clean audio and the targeted transcription for adversarial examples.
\begin{figure}[t]
\includegraphics[width=0.5\textwidth]{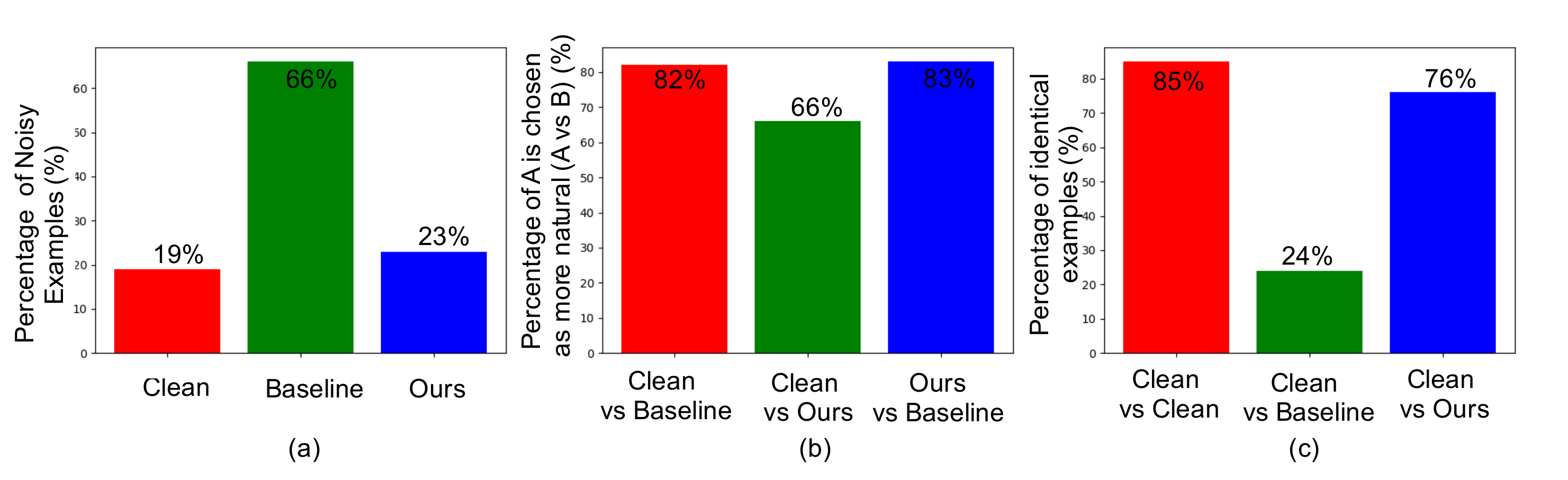}
\vspace{-5mm}
\caption{Results of human study for imperceptibility. Here baseline represents the adversarial example generated by~\citet{Carlini2018AudioAE}, and ours denotes the imperceptible adversarial example generated following the algorithm in Section.~\ref{imperceptible}.}\label{figure}
\vspace{-2mm}
\end{figure}

\subsection{Imperceptibility Analysis}
\label{sec:humanstudy}
To attack the Lingvo ASR system, we construct 1000 imperceptible and targeted adversarial examples, one for each of the examples we sampled from the LibriSpeech test-clean dataset. Table\ref{non_ota_results} shows the performance of the clean audio and the constructed adversarial examples. We can see that the word error rate (WER) of the clean audio is just $4.47\%$ on the 1000 test examples, indicating the model is of high quality. Our imperceptible adversarial examples perform even better, and reach a 100$\%$ success rate.

\subsubsection{Qualitative Human Study}
Of the 1000 examples selected from the test set, we randomly selected $100$ of
these with their corresponding imperceptible adversarial examples. We then generate an adversarial example using the prior work of \citet{Carlini2018AudioAE} for the same target phrase; this attack again succeeds with 100\% success. 
We perform three experiments to validate that our adversarial examples are
imperceptible, especially compared to prior work.

\paragraph{Experimental Design.} We recruit $80$ users
online from Amazon Mechanical Turk.
We give each user one of the three (nearly identical) experiments, each of which we describe below.
In all cases, the experiments consist of 20 ``comparisons tasks'', where we present the evaluator
with some audio samples and ask them questions (described below) about the samples.
We ask the users to listen to each sample with
headphones on, and answer a simple question about the audio samples
(the question is determined by which experiment we run, as given below).
We do not explain the purpose of the study other than that it is a
research study, and do not record any personally identifying information.\footnote{Unfortunately, for this reason, we are unable to report aggregate
statistics such as age or gender, slightly harming potential reproducibility.}
We randomly include a small number of questions with known, obvious answers; 
we remove 3 users from the study who failed to answer these questions
correctly.

In all experiments, users have the ability to listen to audio files
multiple times when they are unsure of the answer, making it as difficult
as possible for our adversarial examples to pass as clean data.
Users additionally have the added benefit of hearing $20$ examples
back-to-back, effectively ``training'' them to recognize subtle differences.
% This is cool but can probably be removed
Indeed, a permutation test finds
users are statistically significantly better at distinguishing
adversarial examples from clean audio during the second half of the
experiment compared to the first half of the experiment, although
the magnitude of the difference is small: only by about $3\%$.
Figure~\ref{figure} summarizes the statistical results we give below.

\paragraph{Experiment 1: clean or noisy.}
We begin with what we believe is the most representative experiment
of how an attack would work in practice. We give users one audio sample and
ask them to tell us if it has \emph{any} background noise (e.g., static,
echoing, people talking in the background).

As a baseline, users believed that $19\%$ of original clean audio samples
contained some amount of noise, and $66\%$ of users believed that
the adversarial examples generated by \citet{Carlini2018AudioAE} contained
some amount of noise.
In comparison, only $23\%$ of users believe that the adversarial examples
we generate contain any noise, a result that is not statistically
significantly different from clean audio ($p>.05$).
That is, when presented with just one audio sample in isolation, users
do not believe the adversarial examples we generate are any noisier
than the clean samples. 

\paragraph{Experiment 2: identify the original.}
We give users two audio samples and inform them that one of the audio
samples is a modified
version of the other; we ask the user to select the audio sample corresponding
to the one which sounds like the \emph{more natural} audio sample.
This setup is much more challenging: when users can listen to both the
before and after, it is often possible to pick up on the small amount
of distortion that has been added.
When comparing the original audio to the adversarial examples generated
by \citet{Carlini2018AudioAE}, the evaluator chose the original audio $82\%$ of
the time.
When we have the evaluator compare the imperceptible adversarial examples we generate to those of \citet{Carlini2018AudioAE},
our imperceptible examples are selected as the better audio samples
$83\%$ of the time---a difference that is not statistically distinguishable
from comparing against the clean audio.

However, when directly comparing the adversarial examples we generate
to the clean audio, users prefer the clean audio still $66\%$ of the time.
Observe that the baseline percentage, when the samples are completely
indistinguishable, is $50\%$. Thus, users only perform $16\%$ better than
random guessing at distinguishing our examples from clean examples.

\paragraph{Experiment 3: identical or not.} Finally, we perform the
most difficult experiment: we present users with two audio files, and ask
them if the audio samples
are identical, or if there are \emph{any} differences.
As the baseline, when given the same audio sample twice, users agreed it was identical $85\%$ of 
the time. (That is, in $15\%$ of cases the evaluator wrongly heard a
difference between the two samples.)
When given a clean audio sample and comparing it to the audio generated
by \citet{Carlini2018AudioAE}, users only believed them to be identical
$24\%$ of the time. 
Comparing clean audio to the adversarial examples we generate, user
believed them to be completely identical $76\%$ of the time, $3\times$ 
more often than the adversarial examples generated by the baseline, but
below the $85\%$-identical value for actually-identical audio.

\subsection{Robustness Analysis} 
To mount our simulated over-the-air attacks, we consider a challenging setting that the exact configuration of the room in which the attack will be performed is unknown. Instead, we are only aware of the distribution from which the room configuration will be drawn. First, we generate 1000 random room configurations sampled from the distribution as the training room set. The test room set includes another 100 random room configurations sampled from the same distribution. Adversarial examples are created to attack the Lingvo ASR system when played in the simulated test rooms. We randomly choose 100 audio examples from LibriSpeech dataset to perform this robustness test.

As shown in Table~\ref{ota_results}, when fed non-adversarial audio played in simulated test rooms, the WER of the Lingvo ASR degrades to $15.42\%$ which suggests some robustness to reverberation. In contrast, the success rate of adversarial examples in~\cite{Carlini2018AudioAE} and our imperceptible adversarial examples in Section~\ref{imperceptible} are $0\%$ in this setting. The success rate of our robust adversarial examples generated based on the algorithm in Section~\ref{secrobust} is over $60\%$, and the WER is smaller than that of the clean audio. Both the success rate and the WER demonstrate that our constructed adversarial examples remain effective when played in the highly-realistic simulated environment. 

In addition, the robustness of the constructed adversarial examples can be improved further at the cost of increased perceptibility. As presented in Table~\ref{ota_results}, when we increase the max-norm bound of the amplitude of the adversarial perturbation $\epsilon^{**}_r = \epsilon^*_r + \Delta$ ($\Delta$ is increased from 300 to 400), both the success rate and WER are improved correspondingly. Since our final objective is to generate imperceptible and robust adversarial examples that can be played over-the-air in the physical world, we limit the max-norm bound of the perturbation to be in a relatively small range to avoid a huge distortion toward the clean audio.

To construct imperceptible as well as robust adversarial examples, we start from the robust attack ($\Delta=300$) and finetune it with the imperceptibility loss. In our experiments, we observe that 81$\%$ of the robust adversarial examples~\footnote{The other 19$\%$ adversarial examples lose the robustness because they cannot successfully attack the ASR system in 8 randomly chosen training rooms in any iteration during optimization.} can be further improved to be much less perceptible while still retaining high robustness (around 50$\%$ success rate and $22.98\%$ WER). 

\subsubsection{Qualitative Human Study}
We run identical experiments (as described earlier) on the robust and robust-and-imperceptible adversarial examples.

In \textbf{experiment 1}, where we ask evaluators if there is any noise, only $6\%$ heard any noise on the clean audio, compared to $100\%$ on the robust (but perceptible) adversarial examples and $83\%$ on the robust and imperceptible adversarial examples. \footnote{Evaluators stated they heard noise on clean examples $3\times$ less often compared to the baseline in the prior study. We believe this is due to the fact that when primed with examples which are obviously different, the baseline becomes more easily distinguishable.}

In \textbf{experiment 2}, where we ask evaluators to identify the original audio, comparing clean to robust adversarial examples the evaluator correctly identified the original audio $97\%$ of the time versus $89\%$ when comparing the clean audio to the imperceptible and robust adversarial examples.

Finally, in \textbf{experiment 3}, where we ask evaluators if the audio is identical, the baseline clean audio was judged different $95\%$ of the time when compared to the robust adversarial examples, and the clean audio was judged different $71\%$ of the time when compared to the imperceptible and robust adversarial examples.

In all cases, the imperceptible and robust adversarial examples are statistically significantly less perceptible than just the robust adversarial examples, but also statistically significantly more perceptible than the clean audio.
Directly comparing the imperceptible and robust adversarial examples to the robust examples, evaluators believed the imperceptible examples had less distortion $91\%$ of the time.

Clearly the adversarial examples that are robust are significantly easier to distinguish from clean audio, even when we apply the masking threshold.
However, this result is consistent with work on adversarial examples on images, where completely imperceptible physical-world adversarial examples have not been successfully constructed. On images, physical attacks require over $16\times$ as much distortion to be effective on the physical world (see, for example, Figure~4 of \citet{kurakin2016adversarial}).

\section{Conclusion}
In this paper, we successfully construct
imperceptible adversarial examples (verified by a human study) for automatic 
speech recognition based on the psychoacoustic principle of auditory masking, 
while retaining 100$\%$ targeted success rate on arbitrary full-sentence targets.  
Simultaneously, we also make progress towards developing robust adversarial examples that remain effective after being played over-the-air (processed by random room environment simulators), increasing the practicality of actual real-world attacks using adversarial examples targeting
ASR systems.

We believe that future work is still required: our robust adversarial examples do not play fully over-the-air, despite working in simulated room environments. Resolving this difficulty while maintaining a high targeted success rate is necessary for demonstrating a practical security concern.

As a final contribution of potentially independent interest, this work demonstrates how one
might go about constructing adversarial examples for non-$\ell_p$-based metrics. Especially on images,
nearly all adversarial example research has focused on this highly-limited distance measure.
Devoting efforts to identifying different methods that humans use to assess similarity, and generating adversarial examples exploiting those metrics, is
an important research effort we hope future work will explore.
% Acknowledgements should only appear in the accepted version.
\section*{Acknowledgements}

The authors would like to thank Patrick Nguyen, Jonathan Shen and Rohit Prabhavalkar for helpful discussions on Lingvo ASR system and Arun Narayanan for suggestions in room impulse simulations. We also want to thank the reviewers for their useful comments.  This work was greatly supported by Google Brain. GWC and YQ were also 
partially supported by Guangzhou Science and Technology
Planning Project (Grant No. 201704030051).

\bibliographystyle{icml2019}

\begin{thebibliography}{29}
\providecommand{\natexlab}[1]{#1}
\providecommand{\url}[1]{\texttt{#1}}
\expandafter\ifx\csname urlstyle\endcsname\relax
  \providecommand{\doi}[1]{doi: #1}\else
  \providecommand{\doi}{doi: \begingroup \urlstyle{rm}\Url}\fi

\bibitem[Allen \& Berkley(1979)Allen and Berkley]{allen1979image}
Allen, J.~B. and Berkley, D.~A.
\newblock Image method for efficiently simulating small-room acoustics.
\newblock \emph{The Journal of the Acoustical Society of America}, 65\penalty0
  (4):\penalty0 943--950, 1979.

\bibitem[Athalye et~al.(2018)Athalye, Engstrom, Ilyas, and
  Kwok]{Athalye2018SynthesizingRA}
Athalye, A., Engstrom, L., Ilyas, A., and Kwok, K.
\newblock Synthesizing robust adversarial examples.
\newblock In \emph{ICML}, 2018.

\bibitem[Bahdanau et~al.(2014)Bahdanau, Cho, and Bengio]{bahdanau2014neural}
Bahdanau, D., Cho, K., and Bengio, Y.
\newblock Neural machine translation by jointly learning to align and
  translate.
\newblock \emph{arXiv preprint arXiv:1409.0473}, 2014.

\bibitem[Biggio et~al.(2013)Biggio, Corona, Maiorca, Nelson, {\v{S}}rndi{\'c},
  Laskov, Giacinto, and Roli]{biggio2013evasion}
Biggio, B., Corona, I., Maiorca, D., Nelson, B., {\v{S}}rndi{\'c}, N., Laskov,
  P., Giacinto, G., and Roli, F.
\newblock Evasion attacks against machine learning at test time.
\newblock In \emph{Joint European conference on machine learning and knowledge
  discovery in databases}, pp.\  387--402. Springer, 2013.

\bibitem[Carlini \& Wagner(2017)Carlini and Wagner]{carlini2017towards}
Carlini, N. and Wagner, D.
\newblock Towards evaluating the robustness of neural networks.
\newblock In \emph{2017 IEEE Symposium on Security and Privacy (SP)}, pp.\
  39--57. IEEE, 2017.

\bibitem[Carlini \& Wagner(2018)Carlini and Wagner]{Carlini2018AudioAE}
Carlini, N. and Wagner, D.~A.
\newblock Audio adversarial examples: Targeted attacks on speech-to-text.
\newblock \emph{2018 IEEE Security and Privacy Workshops (SPW)}, pp.\  1--7,
  2018.

\bibitem[Carlini et~al.(2016)Carlini, Mishra, Vaidya, Zhang, Sherr, Shields,
  Wagner, and Zhou]{carlini2016hidden}
Carlini, N., Mishra, P., Vaidya, T., Zhang, Y., Sherr, M., Shields, C., Wagner,
  D., and Zhou, W.
\newblock Hidden voice commands.
\newblock In \emph{USENIX Security Symposium}, pp.\  513--530, 2016.

\bibitem[Chan et~al.(2016)Chan, Jaitly, Le, and Vinyals]{chan2016listen}
Chan, W., Jaitly, N., Le, Q., and Vinyals, O.
\newblock Listen, attend and spell: A neural network for large vocabulary
  conversational speech recognition.
\newblock In \emph{Acoustics, Speech and Signal Processing (ICASSP), 2016 IEEE
  International Conference on}, pp.\  4960--4964. IEEE, 2016.

\bibitem[Cisse et~al.(2017)Cisse, Adi, Neverova, and Keshet]{cisse2017houdini}
Cisse, M., Adi, Y., Neverova, N., and Keshet, J.
\newblock Houdini: Fooling deep structured prediction models.
\newblock \emph{arXiv preprint arXiv:1707.05373}, 2017.

\bibitem[Gong \& Poellabauer(2017)Gong and Poellabauer]{gong2017crafting}
Gong, Y. and Poellabauer, C.
\newblock Crafting adversarial examples for speech paralinguistics
  applications.
\newblock \emph{arXiv preprint arXiv:1711.03280}, 2017.

\bibitem[Huang et~al.(2017)Huang, Papernot, Goodfellow, Duan, and
  Abbeel]{huang2017adversarial}
Huang, S., Papernot, N., Goodfellow, I., Duan, Y., and Abbeel, P.
\newblock Adversarial attacks on neural network policies.
\newblock \emph{arXiv preprint arXiv:1702.02284}, 2017.

\bibitem[Jia \& Liang(2017)Jia and Liang]{jia2017adversarial}
Jia, R. and Liang, P.
\newblock Adversarial examples for evaluating reading comprehension systems.
\newblock \emph{arXiv preprint arXiv:1707.07328}, 2017.

\bibitem[Khare et~al.(2018)Khare, Aralikatte, and Mani]{khare2018adversarial}
Khare, S., Aralikatte, R., and Mani, S.
\newblock Adversarial black-box attacks for automatic speech recognition
  systems using multi-objective genetic optimization.
\newblock \emph{arXiv preprint arXiv:1811.01312}, 2018.

\bibitem[Kingma \& Ba(2014)Kingma and Ba]{kingma2014adam}
Kingma, D.~P. and Ba, J.
\newblock Adam: A method for stochastic optimization.
\newblock \emph{arXiv preprint arXiv:1412.6980}, 2014.

\bibitem[Kurakin et~al.(2016)Kurakin, Goodfellow, and
  Bengio]{kurakin2016adversarial}
Kurakin, A., Goodfellow, I., and Bengio, S.
\newblock Adversarial examples in the physical world.
\newblock \emph{arXiv preprint arXiv:1607.02533}, 2016.

\bibitem[Lin \& Abdulla(2015)Lin and Abdulla]{lin2015principles}
Lin, Y. and Abdulla, W.~H.
\newblock Principles of psychoacoustics.
\newblock In \emph{Audio Watermark}, pp.\  15--49. Springer, 2015.

\bibitem[Madry et~al.(2017)Madry, Makelov, Schmidt, Tsipras, and
  Vladu]{madry2017towards}
Madry, A., Makelov, A., Schmidt, L., Tsipras, D., and Vladu, A.
\newblock Towards deep learning models resistant to adversarial attacks.
\newblock \emph{arXiv preprint arXiv:1706.06083}, 2017.

\bibitem[Mitchell(2004)]{mitchell2004introduction}
Mitchell, J.~L.
\newblock Introduction to digital audio coding and standards.
\newblock \emph{Journal of Electronic Imaging}, 13\penalty0 (2):\penalty0 399,
  2004.

\bibitem[Panayotov et~al.(2015)Panayotov, Chen, Povey, and
  Khudanpur]{panayotov2015librispeech}
Panayotov, V., Chen, G., Povey, D., and Khudanpur, S.
\newblock Librispeech: an asr corpus based on public domain audio books.
\newblock In \emph{Acoustics, Speech and Signal Processing (ICASSP), 2015 IEEE
  International Conference on}, pp.\  5206--5210. IEEE, 2015.

\bibitem[Scheibler et~al.(2018)Scheibler, Bezzam, and
  Dokmani{\'c}]{scheibler2018pyroomacoustics}
Scheibler, R., Bezzam, E., and Dokmani{\'c}, I.
\newblock Pyroomacoustics: A python package for audio room simulation and array
  processing algorithms.
\newblock In \emph{2018 IEEE International Conference on Acoustics, Speech and
  Signal Processing (ICASSP)}, pp.\  351--355. IEEE, 2018.

\bibitem[Sch{\"o}nherr et~al.(2018)Sch{\"o}nherr, Kohls, Zeiler, Holz, and
  Kolossa]{schonherr2018adversarial}
Sch{\"o}nherr, L., Kohls, K., Zeiler, S., Holz, T., and Kolossa, D.
\newblock Adversarial attacks against automatic speech recognition systems via
  psychoacoustic hiding.
\newblock \emph{arXiv preprint arXiv:1808.05665}, 2018.

\bibitem[Shen et~al.(2019)Shen, Nguyen, Wu, Chen, Chen, Jia, Kannan, Sainath,
  Cao, Chiu, et~al.]{shen2019lingvo}
Shen, J., Nguyen, P., Wu, Y., Chen, Z., Chen, M.~X., Jia, Y., Kannan, A.,
  Sainath, T., Cao, Y., Chiu, C.-C., et~al.
\newblock Lingvo: a modular and scalable framework for sequence-to-sequence
  modeling.
\newblock \emph{arXiv preprint arXiv:1902.08295}, 2019.

\bibitem[Song \& Mittal(2017)Song and Mittal]{song2017inaudible}
Song, L. and Mittal, P.
\newblock Inaudible voice commands.
\newblock \emph{arXiv preprint arXiv:1708.07238}, 2017.

\bibitem[Sutskever et~al.(2014)Sutskever, Vinyals, and
  Le]{sutskever2014sequence}
Sutskever, I., Vinyals, O., and Le, Q.~V.
\newblock Sequence to sequence learning with neural networks.
\newblock In \emph{Advances in Neural Information Processing Systems}, pp.\
  3104--3112, 2014.

\bibitem[Szegedy et~al.(2013)Szegedy, Zaremba, Sutskever, Bruna, Erhan,
  Goodfellow, and Fergus]{szegedy2013intriguing}
Szegedy, C., Zaremba, W., Sutskever, I., Bruna, J., Erhan, D., Goodfellow, I.,
  and Fergus, R.
\newblock Intriguing properties of neural networks.
\newblock \emph{arXiv preprint arXiv:1312.6199}, 2013.

\bibitem[Taori et~al.(2018)Taori, Kamsetty, Chu, and Vemuri]{taori2018targeted}
Taori, R., Kamsetty, A., Chu, B., and Vemuri, N.
\newblock Targeted adversarial examples for black box audio systems.
\newblock \emph{arXiv preprint arXiv:1805.07820}, 2018.

\bibitem[Yakura \& Sakuma(2018)Yakura and Sakuma]{yakura2018robust}
Yakura, H. and Sakuma, J.
\newblock Robust audio adversarial example for a physical attack.
\newblock \emph{arXiv preprint arXiv:1810.11793}, 2018.

\bibitem[Yuan et~al.(2018)Yuan, Chen, Zhao, Long, Liu, Chen, Zhang, Huang,
  Wang, and Gunter]{yuan2018commandersong}
Yuan, X., Chen, Y., Zhao, Y., Long, Y., Liu, X., Chen, K., Zhang, S., Huang,
  H., Wang, X., and Gunter, C.~A.
\newblock Commandersong: A systematic approach for practical adversarial voice
  recognition.
\newblock \emph{arXiv preprint arXiv:1801.08535}, 2018.

\bibitem[Zhang et~al.(2017)Zhang, Yan, Ji, Zhang, Zhang, and
  Xu]{zhang2017dolphinattack}
Zhang, G., Yan, C., Ji, X., Zhang, T., Zhang, T., and Xu, W.
\newblock Dolphinattack: Inaudible voice commands.
\newblock In \emph{Proceedings of the 2017 ACM SIGSAC Conference on Computer
  and Communications Security}, pp.\  103--117. ACM, 2017.

\end{thebibliography}

\section*{Appendix}
\appendix
\section{Frequency Masking Threshold}
In this section, we detail how we compute the frequency masking threshold for constructing imperceptible adversarial examples. This procedure is based on psychoacoustic principles which were refined over many years of human studies. For further background on psychoacoustic models, we refer the interested reader to~\cite{lin2015principles, mitchell2004introduction}.

\textbf{Step 1: Identifications of Maskers}

\vspace{1mm}
In order to compute the frequency masking threshold of an input signal $x(n)$, where $0\leq n \leq N$, we need to first identify the maskers. There are two different classes of maskers: tonal and nontonal maskers, where nontonal maskers have stronger masking effects compared to tonal maskers. Here we simply treat all the maskers as tonal ones to make sure the threshold that we compute can always mask out the noise. The normalized PSD estimate of the tonal maskers $\bar{p}^m_x(k)$ must meet three criteria. First, they must be local maxima in the spectrum, satisfying:
\begin{equation}
    \bar{p}_x(k - 1) \leq \bar{p}_x^m (k) \quad \text{and} \quad \bar{p}_x^m (k)\geq \bar{p}_x(k+1), 
\end{equation}
where $0\leq k < \frac{N}{2}$.

Second, the normalized PSD estimate of any masker must be higher than the threshold in quiet ATH($k$), which is:
\begin{equation}
    \bar{p}^m_x(k)  \geq \text{ATH} (k),
\end{equation}
where ATH($k$) is approximated by the following frequency-dependency function:
\begin{equation} \label{lnet}
\begin{split}
\text{ATH}(f) & = 3.64 (\frac{f}{1000})^{-0.8} - 6.5\exp \{ -0.6(\frac{f}{1000} - 3.3)^2\}\\
 & \quad \quad + 10^{-3}(\frac{f}{1000})^4.
\end{split}
\end{equation}
The quiet threshold only applies to the human hearing range of $20\text{Hz} \leq f \leq 20 \text{kHz}$. When we perform short time Fourier transform (STFT) to a signal, the relation between the frequency $f$ and the index of sampling points $k$ is 
\begin{equation}
f = \frac{k}{N} \cdot f_s,   \quad 0\leq f < \frac{f_s}{2}
\end{equation}
where $f_s$ is the sampling frequency and $N$ is the window size.

Last, the maskers must have the highest PSD within the range of 0.5 Bark around the masker's frequency, where bark is a psychoacoustically-motivated frequency scale. Human's main hearing range between 20Hz and 16kHz is divided into 24 non-overlapping critical bands, whose unit is Bark, varying as a function of frequency $f$ as follows:
\begin{equation}
    b(f) = 13\arctan (\frac{0.76f}{1000}) + 3.5 \arctan (\frac{f}{7500})^2.
\end{equation}

As the effect of masking is additive in the logarithmic domain, the PSD estimate of the the masker is further smoothed with its neighbors by:
\begin{equation}
    \bar{p}^m_x(\bar{k}) = 10 \log_{10} [10^{\frac{\bar{p}_x(k-1)}{10}} + 10^{\frac{\bar{p}^m_x(k)}{10}} + 10^{\frac{\bar{p}_x(k+1)}{10}} ]
\end{equation}
\textbf{Step 2: Individual masking thresholds}

\vspace{2mm}
An individual masking threshold is better computed with frequency denoted at the Bark scale because the spreading functions of the masker would be similar at different Barks. We use $b(i)$ to represent the bark scale of the frequency index $i$. There are a number of spreading functions introduced to imitate the characteristics of maskers and here we choose the simple two-slope spread function:

\begin{equation}
\text{SF}[b(i), b(j)] =\begin{cases}
     27 \Delta b_{ij}, \quad \text{if $\Delta b_{ij} \leq 0$}.\\
      G(b(i)) \cdot \Delta b_{ij}, \quad \text{otherwise}\\
  
  \end{cases}
\end{equation}

where 
\begin{equation}
\Delta b_{ij} = b(j) - b(i),
\end{equation}
\begin{equation}
G(b(i)) = [-27+0.37\max\{\bar{p}^m_x(b(i)) -40, 0\}]
\end{equation}

where $b(i)$ and $b(j)$ are the bark scale of the masker at the frequency index $i$ and the maskee at frequency index $j$ respectively. Then, $T[b(i),b(j)]$ refers to the masker at Bark index $b(i)$ contributing to the masking effect on the maskee at bark index $b(j)$. Empirically, the threshold $T[b(i),b(j)]$ is calculated by:
\begin{equation}
     T[b(i),b(j)] = \bar{p}^m_x(b(i)) + \Delta_m[b(i)] + \text{SF}[b(i), b(j)],
\end{equation}
where $\Delta_m[b(i)] = -6.025 -0.275b(i)$ and SF[$b(i), b(j)$] is the spreading function. 
\vspace{4mm}\\
\textbf{Step 3: Global masking threshold}

\vspace{2mm}
The global masking threshold is a combination of individual masking thresholds as well as the threshold in quiet via addition. The global masking threshold at frequency index $i$ measured with Decibels (dB) is calculated according to:
\begin{equation}
    \theta_x(i) = 10 \log_{10}[10^{\frac{ATH(i)}{10}} + \sum_{j=1}^{N_m} 10^{\frac{T[b(j), b[i]]}{10}}],
\end{equation}
 where $N_m$ is the number of all the selected maskers. The computed $\theta_x$ is used as the frequency masking threshold for the input audio $x$ to construct imperceptible adversarial examples.

\section{Stability in Optimization} In case of the instability problem during back-propagation due to the existence of the $\log$ function in the threshold $\theta_x(k)$ and the normalized PSD estimate of the perturbation $\bar{p}_{\delta}(k)$, we remove the term $10\log_{10}$ in the PSD estimate of $p_{\delta}(k)$ and $p_x(k)$ and then they become:
\begin{equation}
    p_{\delta}(k) = \left|\frac{1}{N}s_{\delta}(k)\right|^2, \quad p_{x}(k) = \left|\frac{1}{N}s_x(k)\right|^2
\end{equation}
and the normalized PSD of the perturbation turns into
\begin{equation}
    \bar{p}_{\delta}(k) =\frac{10^{9.6}p_{\delta}(k)}{\max_k\{ p_x(k)\}}.
\end{equation}
Correspondingly, the threshold $\theta_x(k)$ becomes:
\begin{equation}
    \theta_x(k) = {10}^{\theta_x \over 10}%10^{\frac{th_x(k)}{10}}
\end{equation}

\section{Notations and Definitions}
The notations and definitions used in our proposed algorithms are listed in Table~\ref{notation}.
 \begin{table*}[]
\centering
\begin{tabular}{l|l}
\hline
$x$ & The clean audio input \\ \hline
 $\delta$ & The adversarial perturbation added to clean audio                   \\ \hline
 $x'$ & The constructed adversarial example                  \\ \hline
 $y$ & The targeted transcription                 \\ \hline
 $f(\cdot)$ & The attacked neural network (ASR)                  \\ \hline
 $\mathscr{F}(\cdot)$ & Fourier transform                 \\ \hline
 $k$ & The index of the spectrum              \\ \hline
 $N$ & The window size in short term Fourier transform              \\ \hline
 $s_x(k)$ & The $k$-th bin of the spectrum for audio $x $              \\ \hline
 $s_\delta(k)$ & The $k$-th bin of the spectrum for perturbation $\delta$               \\ \hline
 $p_x(k)$ & The log-magnitude power spectral density (PSD) for audio $x$ at index $k$              \\ \hline
 $\bar{p}_x(k)$ & The normalized PSD estimated for audio $x$ at index $k$             \\ \hline
 $p_\delta(k)$ & The log-magnitude power spectral density (PSD) for audio $\delta$ at index $k$            \\ \hline
 $\bar{p}_\delta(k)$ & The normalized PSD estimated for audio $\delta$ at index $k$             \\ \hline
 $\theta_x(k)$ & The frequency masking threshold for audio $x$ at index $k$           \\ \hline

 $\ell(x, \delta, y)$ & Loss function to optimize to construct adversarial examples         \\ \hline
 $\ell_{net}(\cdot, y)$ & Loss function to fool the neural network with the input $(\cdot)$ and output $y$        \\ \hline
 $\ell_\theta(x, \delta)$ & Imperceptibility loss function          \\ \hline
 $\alpha$ & A hyperparameter to balance the importance of $\ell_{net}$ and $\ell_\theta$        \\ \hline
 $\lVert{} \cdot \rVert{}$ & Max-norm         \\ \hline
 $\epsilon$ & Max-norm bound of perturbation $\delta$         \\ \hline
 $\nabla_\delta (\cdot)$ & The gradient of ($\cdot$) with regard to $\delta$        \\ \hline
$lr_1, lr_2, lr_3$ & The learning rate in gradient descent       \\ \hline
$r$ & Room reverberation       \\ \hline
$t(\cdot)$ & The room transformation related to the room configuration       \\ \hline
$\mathcal{T}$ & The distribtion from which the transformation $t(\cdot)$ is sampled from       \\ \hline
$\delta^*_{im}$ & The optimized $\delta$ in the first stage in constructing imperceptible adversarial examples     \\ \hline
$\epsilon^*_{r}$ & The optimized $\epsilon$ in the first stage in constructing robust adversarial examples   \\ \hline
$\delta^*_{r}$ & The optimized $\delta$ in the first stage in constructing robust adversarial examples     \\ \hline
$\epsilon^{**}_{r}$ & The max-norm bound for $\delta$ used in the second stage in constructing robust adversarial examples     \\ \hline
$\delta^{**}_{r}$ & The optimized $\delta$ in the second stage in constructing robust adversarial examples     \\ \hline
$\Delta$ & The difference between $\epsilon^{**}_r - \epsilon^*_r$     \\ \hline
$\Omega$ & A set of transformations sampled from distribution $\mathcal{T}$  \\ \hline
$M$ & The size of the transformation set $\Omega$     \\ \hline
\end{tabular}
\caption{Notations and Definitions used in our algorithms.}\label{notation}
\end{table*}

\section{Implementation Details}
The adversarial examples generated in our paper are all optimized via Adam optimizer~\cite{kingma2014adam}. The hyperparameters used in each section are displayed below.
\subsection{Imperceptible Adversarial Examples}
In order to construct imperceptible adversarial examples, we divide the optimization into two stages. In the first stage, the learning rate $lr_1$ is set to be 100 and the number of iterations $T_1$ is 1000 as~\cite{Carlini2018AudioAE}. The max-norm bound $\epsilon$ starts from 2000 and will be gradually reduced during optimization. In the second stage, the number of iterations $T_2$ is 4000. The learning rate $lr_2$ starts from 1 and will be reduced to be 0.1 after 3000 iterations. The adaptive parameter $\alpha$ which balances the importance between $\ell_{net}$ and $\ell_\theta$ begins with $0.05$ and gradually updated based on the performance of adversarial examples. Algorithm~\ref{im} shows the details of the two-stage optimization.

\begin{algorithm}[tb]
   \caption{Optimization with Masking Threshold}
   \label{im}
\begin{algorithmic}
   \STATE {\bfseries Input:} audio waveform $x$, target phrase $y$, ASR system $f(\cdot)$, perturbation $\delta$, loss function $\ell(x, \delta, y)$, hyperparameters $\epsilon$ and $\alpha$, learning rate in the first stage $lr_{1}$ and second stage $lr_{2}$, number of iterations in the first stage $T_1$ and second stage $T_2$.
   \STATE \# Stage 1: minimize $\lVert{}\delta\rVert{}$
   \STATE Initialize $\delta = 0$, $\epsilon=2000$ and $\alpha=0$.
   \FOR{$i=0$ {\bfseries to} $T_1-1$}
  
   \STATE $\delta \leftarrow \delta - lr_{1} \cdot \textnormal{sign}(\nabla_\delta \ell(x, \delta, y) ) $
   \STATE{Clip $\lVert{}\delta\rVert{} \leq \epsilon$}
   \IF{$i$  $\%$ $10 = 0 $ and $f(x+\delta) = y$}
   \IF{$\epsilon > \max(\lVert{}\delta\rVert{})$}
   \STATE $\epsilon \leftarrow {\max(\lVert{}\delta\rVert{})}$
   \ENDIF
   \STATE $\epsilon\leftarrow 0.8 \cdot \epsilon $
   \ENDIF
   \ENDFOR
   \STATE{}
   \STATE{\# Stage 2: minimize the perceptibility}
   \STATE{Reassign $\alpha = 0.05$}
   \FOR{$i=0$ {\bfseries to} $T_2-1$}
  
   \STATE $\delta \leftarrow \delta - lr_{2} \cdot \nabla_\delta \ell(x, \delta, y)$
   \IF{$i$  $\%$ $20 = 0 $ and $f(x+\delta) = y$}
   \STATE $\alpha\leftarrow 1.2 \cdot \alpha $
   \ENDIF
   \IF{$i$  $\%$ $50 = 0 $ and $f(x+\delta) \neq y$}
   \STATE $\alpha\leftarrow 0.8 \cdot \alpha $
   \ENDIF
   \ENDFOR
   \STATE {\bfseries Output:} adversarial example $x' = x + \delta$
\end{algorithmic}
\end{algorithm}
\begin{table*}[]
\centering
\begin{tabular}{l|l}
\hline
\hline
Original phrase 1 & the more she is engaged in her proper duties the less leisure will she have for it even as an \\ & accomplishment and a recreation \\ \hline
Targeted phrase 1 & old will is a fine fellow but poor and helpless since missus rogers had her accident                                          \\ \hline\hline
Original phrase 2 & a little cracked that in the popular phrase was my impression of the stranger who now made his \\ & appearance in the supper room  \\ \hline
Targeted phrase 2 & her regard shifted to the green stalks and leaves again and she started to move away                                          \\ \hline\hline
\end{tabular}
\caption{Examples of the original and targeted phrases on the LibriSpeech dataset.}\label{example}
\end{table*}
\subsection{Robust Adversarial Examples}\label{robust}
To develop the robust adversarial examples that could work after played over-the-air, we also optimize the adversarial perturbation in two stages. The first stage intends to find a relative small perturbation while the second stage focuses on making the constructed adversarial example more robust to random room configurations. The learning rate $lr_1$ in the first stage is $50$ and $\delta$ will be updated for 2000 iterations. The max-norm bound $\epsilon$ for the adversarial perturbation $\delta$ starts from 2000 as well and will be gradually reduced. In the second stage, the number of iterations is set to be 4000 and the learning rate $lr_2$ is $5$. In this stage, $\epsilon$ is fixed and equals the optimized $\epsilon^*_r$ in the first stage plus $\Delta$. The size of transformation set $\Omega$ is set to be $M=10$.
\subsection{Imperceptible and Robust Attacks}
To construct imperceptible as well as robust adversarial examples, we begin with the robust adversarial examples generated in Section.~\ref{robust}. In the first stage, we focus on reducing the imperceptibility by setting the initial $\alpha$ to be 0.01 and the learning rate is set to be 1. We update the adversarial perturbation $\delta$ for 4000 iterations. If the adversarial example successfully attacks the ASR system in 4 out of 10 randomly chosen rooms, then $\alpha$ with be increased by 2. Otherwise, for every 50 iterations, $\alpha$ will be decreased by 0.5.

In the second stage, we focus on improving the less perceptible adversarial examples to be more robust. The learning rate is 1.5 and $\alpha$ starts from a very small value $0.00005$. The perturbation will be further updated for 6000 iterations. If the adversarial example successfully attacks the ASR system in 8 out of 10 randomly chosen rooms, then $\alpha$ will be increased by 1.2.

\section{Transcription Examples}
Some examples of the original phrases and targeted transcriptions from the LibriSpeech dataset~\cite{panayotov2015librispeech} are shown in Table~\ref{example}.

\end{document}